\documentclass[runningheads]{llncs}
\usepackage{graphicx}
\usepackage{todonotes}

\begin{document}
\title{Human Factors in Space Exploration: Opportunities for International and Interdisciplinary Collaboration}
\titlerunning{Human Factors in Space Exploration Panel}
%
\author{Wiesław Kopeć \inst{1,4}\orcidID{0000-0001-9132-4171} \and
Grzegorz Pochwatko\inst{2}\orcidID{0000-0001-8548-6916} \and
Monika Kornacka\inst{3}\orcidID{0000-0003-2737-9236} \and
Wiktor Stawski \inst{1}\orcidID{0000-0001-8950-195X} \and
Maciej Grzeszczuk\inst{1}\orcidID{0000-0002-9840-3398} \and
Kinga Skorupska\inst{1}\orcidID{0000-0002-9005-0348} \and
Barbara Karpowicz\inst{1}\orcidID{0000-0002-7478-7374} \and
Rafał Masłyk\inst{1}\orcidID{0000-0003-1180-2159} \and
Pavlo Zinevych \inst{1} \orcidID{0009-0008-9250-8712} \and
Stanisław Knapiński \inst{1}\orcidID{0009-0006-0524-2545} \and
Steven Barnes\inst{3}\orcidID{0000-0002-5114-2178}  \and
Cezary Biele\inst{5}\orcidID{0000-0003-4658-5510} 
}
\authorrunning{Kopec et al.}
%
\institute{XR Center, Polish-Japanese Academy of Information Technology
 \url{https://xrc.pja.edu.pl}
\and
Institute of Psychology, Polish Academy of Sciences
\and
SWPS University of Social Sciences and Humanities 
\and
Kobo Association
\and National Information Processing Institute
}

\maketitle              
\begin{abstract}

As humanity pushes the boundaries of space exploration, human factors research becomes more important. Human factors encompass a broad spectrum of psychological, physiological, and ergonomic factors that affect human performance, well-being, and safety in the unique and challenging space environment. This panel explores the multifaceted field of human factors in space exploration and highlights the opportunities that lie in fostering international and interdisciplinary cooperation. This exploration delves into the current state of research on human factors in space missions, addressing the physiological and psychological challenges astronauts face during long space flights. It emphasizes the importance of interdisciplinary collaboration, combining knowledge from fields such as psychology, medicine, engineering, and design to address the complex interaction of factors affecting human performance and adaptation to the space environment.


\end{abstract}
\section{Rationale}

This panel is a follow-up of last year's talk at the opening session of the 10th anniversary MIDI conference. Last year was also the 10th anniversary of Polish membership in the European Space Agency. Therefore, during the special panel \textit{Challenges and Opportunities in Space Exploration using eXtended Reality} we opened discussion within the broader group of researchers and professionals, including representatives from ESA EAC, XR Lab, Polish Space Agency (POLSA), Polish Ministry of Sciences and Education, Space Research Center of Polish Academy of Sciences, and the APLHA-XR research framework team of HASE (Human Aspects of Science and Engineering) research network by Living Lab Kobo, including the VR Lab Institute of Psychology Polish Academy of Sciences, EC Lab SWPS University, LIT National Information Processing Institute, and XR Center Polish Academy of Information Technologies. 

\begin{figure*}[ht!]
 \centering
 \includegraphics[width=1\textwidth]{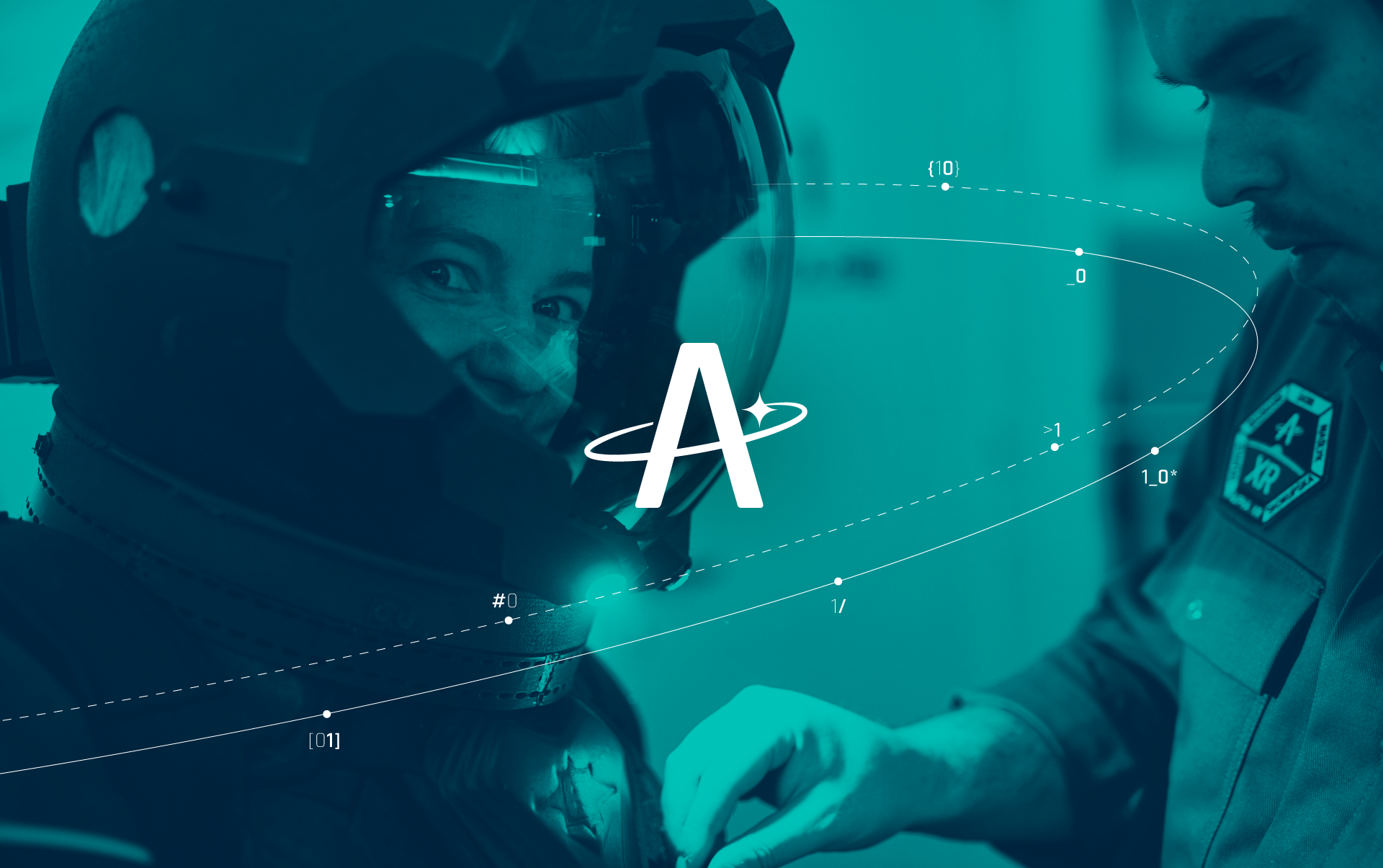}
 \caption{ALPHA-XR Simulated ICE Study. \\Photo by Suzanne Baumann, collage by Mateusz Szewczyk, Art\&Design Lab, XRC}
 \label{fig:alphaxr}
\end{figure*}

This time we would like to focus on certain aspects of human factors in space exploration based on opportunities and challenges that stem from our recent research endeavors and plans within ALPHA-XR initiative. This initiative employs a multimodal data-driven approach utilising new technologies such as Augmented and Virtual Reality (AR, VR) which comprise the eXtended Reality (XR) continuum. 

During last year's discussion it was clear that space exploration, as one of the main pillars of space research and in addition to Earth observation, is also one of the most challenging and exciting areas of research in general. This is mainly due to the strong interdisciplinarity and potential toward practical implementation beyond the space research area. Its importance is reflected in the strategies of space agencies, including NASA and ESA. NASA's vision is to \textit{explore the secrets of the universe for the benefit of all}\footnote{\url{https://www.nasa.gov/humans-in-space/why-go-to-space/}}, and according to its mission, NASA explores \textit{the unknown in air and space, innovates for the benefit of humanity, and inspires the world through discovery}\footnote{\url{https://www.nasa.gov/about/}}. 

The importance of space exploration is also reflected in the strategy of the European Space Agency (ESA). In particular, the 2014 \textit{Resolution on Europe’s space exploration} strategy adopted at the Ministerial level of the ESA Council paved the way for the evolution of successful space cooperation in Europe. It is based on more than half a century of previous experience in the field to maintain Europe (and the ESA in particular) as one of the world leaders within the area of space science, Earth observation, space exploration, and related technology. Based on that, \textit{The European Exploration Envelope Programme}(E3P) was created in 2016 to implement the European space exploration strategy. The E3P brought together all ESA exploration activities in a single program. To underline the importance of space exploration and European ambitions in this area, the E3P program was rebranded in 2021 as \textit{Terrae Novae} (New Worlds) and followed by \textit{Terrae Novae 2030+ Strategy Roadmap}\footnote{$esamultimedia.esa.int/docs/HRE/Terrae\_Novae\_2030+strategy\_roadmap.pdf$} released in 2022. According to the official documentation of the strategy, \textit{the mission of the Terrae Novae program is to lead the human journey of Europe to the solar system using robots as precursors and scouts, and to return the benefits of exploration back to society}. As an ambitious exploration vision for Europe, it comprises of threefold objectives related to \textbf{Low Earth Orbit} (LEO), \textbf{Moon} and \textbf{Mars}, namely:
\begin{itemize}
    \item \textit{to create new opportunities in low Earth Orbit for a sustained European presence after the International Space Station}, 
    \item \textit{to enable the first European to explore the Moon’s surface by 2030 as a step towards sustainable lunar exploration in the 2030’s}, and
    \item \textit{to prepare the horizon goal of Europe being part of the first human mission to Mars}.
\end{itemize}

According to the official documentation, this strategic vision \textit{evokes the spirit of new discoveries, new ambitions, new science, new inspiration, and new challenges. It symbolizes the constant quest for technological, process, and procurement innovations that result in new and better ways to deliver the program. It also reflects the aspiration to actively reach out to new partners beyond the space sector and expand the space ecosystem to the commercial sphere}.

Having in mind strategic plans at the international level, as well as the growing importance of space research in Poland, in 2022 we established a proposal towards a sustainable XR-based Space Exploration Research Framework called ALPHA-XR. It serves as a new long-term cooperation framework for transdisciplinary research that fits in the Strategic Innovation Area of Human and Robotic Exploration, part of Terrae Novae program. Based on our previous experience as a HASE research initiative, the primary goal of ALPHA-XR was to develop and utilise a data-driven approach to immersive research environments for evidence-based transdisciplinary studies in ICE (Isolation, Confinement, Extreme) conditions, with broad field of scientific and practical applications including:
\begin{itemize}
    \item XR-based research in ICE condition, 
    \item XR-based interventions and training in ICE condition, 
    \item XR-supported communication and cooperation including robotic interactions. 
\end{itemize}

\begin{figure*}[ht!]
 \centering
 \includegraphics[width=1\textwidth]{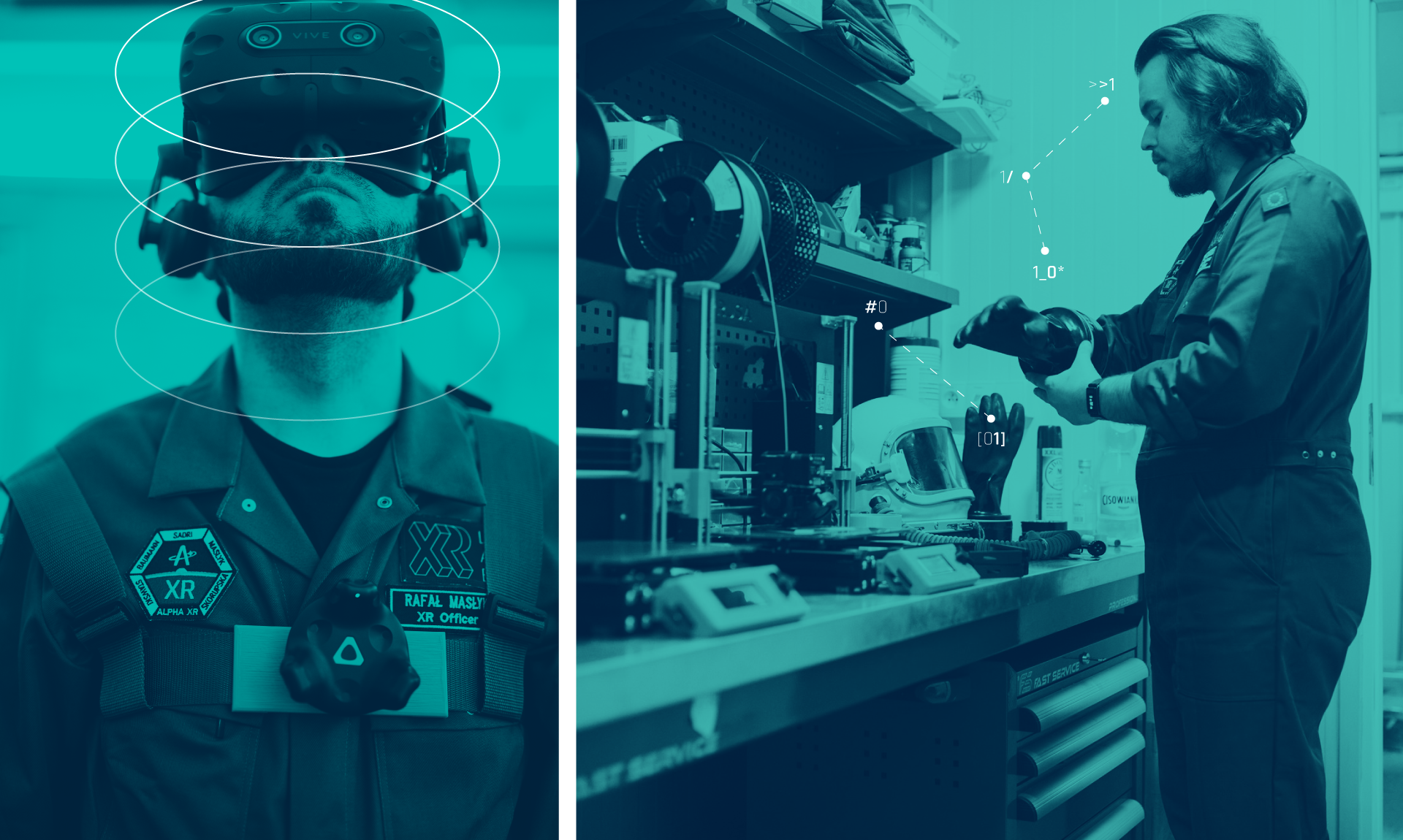}
 \caption{ALPHA-XR Simulated ICE Study, VR Session \&3D printing prototyping. \\Photo by Suzanne Baumann, collage by Mateusz Szewczyk, Art\&Design Lab, XRC}
 \label{fig:alphaxr2}
\end{figure*}

The assumptions of the APLHA-XR framework (alpha.xrlab.pl) were presented and discussed with representatives of POLSA and the Polish Ministry of Science as well as other collaborating institutions, including the ESA EAC, XR Lab, and nvLab Lodz Film School and served also as a canvas for last year’s MIDI 2022 opening panel discussion. The above-mentioned areas of ALPHA-XR were also a foundation of our contribution to the ESA Topical Team on Space Analogs and Human Performance. Based on the proposal during the last two years, we have conducted some exploratory research and studies, including manned space simulations (fig. \ref{fig:alphaxr}) using mixed research methods and multimodal data that included psychophysiological data coupled with data from validated questionnaires and ecological momentary assessment acquired in controlled, ecologically valid conditions supported by the use of the mixed reality continuum, including real and virtual immersive environments (fig. \ref{fig:alphaxr2}). As we mentioned before, the current MIDI 2023 discussion panel is a follow-up of the open discussion on human factors in space exploration that we started last year at the conference.

\section{Theme}

During the panel, we will discuss the recent advancements of our research collaboration network on human factors in space exploration that stem from our ALPHA-XR framework proposal. According to the proposal, the research activities were conducted by HASE group core labs (XRC PJAIT, EC Lab USWPS, VR Lab IP PAS, and LIT NIPI) in collaboration with other teams and institutions, including ESA EAC XR Lab, nvLab Lodz Film School, as well as ESA Topical Team Space Analogs and Human Performance.

\subsection{Focus}

The discussion will focus on both the positive results of the research activities (which have been partially published) and the identified challenges that were also part of our contribution to the ESA Topical Team. In particular, the discussion points refer to the crucial elements of the ALPHA-XR program, namely, results and challenges related to conducting advanced studies and application based on multimodal data-driven immersive environments as part of an evidence-based multidisciplinary and multimodal psychological foundation to the main research topics of ALPHA-XR:
\begin{itemize}
    \item identification and detection of key factors that influence emotional regulation and well-being affecting also executive and attentional functions,
    \item providing effective ways of monitoring emotional regulation and well-being in order to maintain their optimal level, and prevent lowering individual and crew skills, executive performance and well-being,
    \item supporting human attention, perception and just-in-time learning in ICE conditions with XR solutions, based on the aforementioned monitoring methods.
\end{itemize}

\subsection{Discussion Points}
Application of the recent capabilities of immersive data-driven environments in the context of human factors in space exploration is, according to our ALPHA-XR program, one of the promising yet challenging areas of research and development. Therefore, we will discuss several aspects related to the application of the data-driven XR-based space exploration research which are bridging the gap between regular experimental laboratory studies and field studies.

\paragraph{Aspect 1: New opportunities for space exploration}
Although there is a large body of knowledge on human factors related to space exploration, there is still a need for further research in isolated conditions, mainly due to the specific conditions and requirements of planned manned missions to the Moon and, further, to Mars. Besides medical challenges which have been well recognized in both ground-based facilities as well as natural space conditions, such as exposure to microgravity or cosmic radiation, prolonged isolation, which may also significantly affect crew performance and well-being, needs further research, methods and tools. One well-established concept is simulated missions in analogue habitats to provide better ecological validity.

\paragraph{Aspect 2: Limitations of current analog research facilities}
Even though the analog habitats and missions are well established in research in terms of ecological validity of isolation and confinement research aspects, there are certain limitations that stem from various factors of isolation studies: from challenges with conducting proper measurement in unsupervised conditions to small study samples and one-shot studies, as well as differences in equipment, setting and mission routines that lead to difficulties in study replication.

\paragraph{Aspect 3: Bridging the gaps in isolation and confinement research}
Nevertheless, thanks to recent research and endeavours, we contribute some positive results of our exploratory studies in ecologically valid analog missions in terms of various aspects of research, including intervention, communication or prototyping.

\section{Organizers}

\subsection{Panelists}

\paragraph{Wiesław Kopeć}
Computer scientist, research and innovation team leader, associate professor at Computer Science Faculty of Polish-Japanese Academy of Information Technology (PJAIT). Head of XR Center PJAIT and XR Department. He is also a seasoned project manager with a long-term collaboration track with many universities and academic centers, including University of Warsaw, SWPS University, National Information Processing Institute, and institutes of Polish Academy of Sciences. He co-founded the transdisciplinary HASE research group (Human Aspects in Science and Engineering) and distributed LivingLab Kobo.

\paragraph{Grzegorz Pochwatko}
Psychologist, researcher, and academic teacher. He has established and currently leads the Virtual Reality and Psychophysiology Lab at the Institute of Psychology of the Polish Academy of Sciences. His research interests focus on the behaviors of users within virtual reality environments. In particular, he deals with interactions between avatars, humanoid virtual agents, and robots. Specifically, his expertise lies in analyzing the complex interactions among avatars, humanoid virtual agents, and robots. Additionally, his work extensively covers the phenomenon of co-presence, examining its nuances in both computer-generated virtual environments and the immersive realm of cinematic VR. 

\paragraph{Monika Kornacka}
Psychologist and researcher. Her research interests focus on emotion regulation, transdiagnostic processes (especially repetitive negative thoughts and divagation of thought), and cognitive-behavioral therapy. She is also interested in executive function training and the application of new technologies in clinical psychology and in research on psychopathology. She is a head of the Emotion Cognition Lab (ECL), where she uses the latest methods and technical solutions, such as mobile applications, eyetracking, virtual and extended reality, to measure psychological processes, and test how new technologies may be applied to psychological interventions.

\paragraph{Wiktor Stawski}
Enthusiast of Space Research, 3D modeling and printing. Young researcher of XR Center PJAIT, co-founder and head of XR Students' Club. He gained first-hand experience of human factors in space exploration as a crew member of ALPHA-XR simulated two-week mission in lunar habitat. He has also published the results of his exploratory study on the largest human-computer conference CHI 2023\cite{stawski2023druk}.

\paragraph{Maciej Grzeszczuk}
Data scientist with a diverse professional background, including telecommunications, business processes, aviation, and air traffic control. Member of the Department of XR and Immersive Systems at the Polish-Japanese Academy of Information Technology, researching issues related to the human factor in extreme conditions and archiving old magnetic media in order to preserve cultural heritage. Privately, a fan of retrocomputers, sometimes a pilot, and more often a sailor and traveler.

\begin{figure*}[ht!]
 \centering
 \includegraphics[width=1\textwidth]{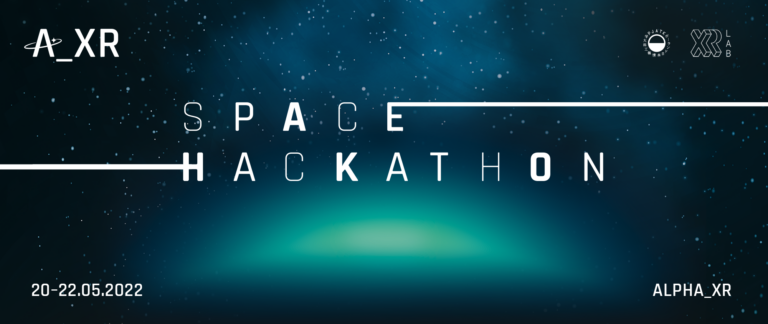}
 \caption{ALPHA-XR Space Hackathon.\\ Graphic design by Mateusz Szewczyk, Art\&Design Lab, XR Center PJAIT}
 \label{fig:xrhackathon}
\end{figure*}

\subsection{Researchers network}
We would like to thank the many people and institutions, gathered by the Living Lab Kobo and the HASE Research Group, to allow for this collaboration and research. Firstly, we thank all the members of HASE research group (Human Aspects in Science and Engineering) and Living Lab Kobo for their support. In particular, the members of the XR Center Polish-Japanese Academy of Information Technology (XRC PJAIT), including Mateusz Szewczyk, head of XRC Art\&Design Lab for visual identification system, as well as its students' club, supporting ALPHA-XR efforts, in particular Wiktor Stawski and Stanisław Knapiński as well as XR Center Staff involved in this process, that is Barbara Karpowicz, Rafał Masłyk and Pavlo Zinevych, and the Emotion-Cognition Lab at SWPS University (EC Lab), Kobo Association, Living Lab Kobo community, VR and Psychophisiology Lab of the Institute of Psychology Polish Academy of Sciences, and Laboratory of Interactive Technologies (LIT) National Information Processing Institute (NIPI). We would also like to thank Paulina Borkiewicz from Visual Narratives Laboratory, Lodz Film School, and Agnieszka Kuczała from ESA Programme Board for Human Spaceflight, Microgravity and Exploration, as well as the members of ESA Topical Team Space Analogs and Human Performance led by Gabriel G. de la Torre, and XR Lab ESA EAC, in particular Tommy Nilsson for their support of the joint efforts on exploring human factors in space exploration in VR and beyond.

\section{Discussion and Conclusions}
Our panel discussion last year took place the day after the successful landing of the Orion module on the Artemis 1 space mission - one of the cornerstones of returning humans to the Moon more than half a century since the last Apollo landing. This was an example of direct evidence that the above-mentioned strategic plans and roadmaps are not just theoretical outlines repeatedly postponed in time, but rather a real frame of reference for international and interdisciplinary cooperation that may lead to reestablish human presence on the Moon in upcoming years. Therefore, various efforts towards strengthening the capacity of human factors research, intervention, and training aspects of space exploration become even more relevant. It is even more important than decades ago in the Apollo era, when crew members had military service backgrounds, predominantly as former pilots. On the contrary, in the Artemis program, the crew consists mainly of scientists or professionals without years of special military training or experience, making matters related to long-term isolated manned space missions even more challenging than those of decades ago.

Although there is a tremendous body of knowledge gathered in various fields and disciplines concerning different aspects related to space missions, there is still room for new research and bridging the gap in interdisciplinary collaboration. Therefore, we are pleased that our recent endeavors within the ALPHA-XR research framework allowed us to organize a simulated isolation analog lunar mission (fig. \ref{fig:alphaxr}). This allowed us to conduct a series of studies and subsequent research activities that shed additional light on some aspects related to human performance in isolation related to well-being \cite{ismar2023asthenia}, effective communication \cite{skorupska2022case}{} and rapid prototyping \cite{stawski2023druk} as well as other related aspects such as conducting unsupervised XR-based studies \cite{interact2023kopec}.

For better understanding of our recent research findings and publications, including promising application of IVR, as well as certain limitations and shortcomings of analog habitat missions, it is important to bring some context to our discussion of isolation studies. In particular, why they are performed in the specific way and what the implications are, both positive (that should be the subject of further development) and negative (that might or should be mitigated).

As mentioned above, one of the crucial challenges of human factors in space exploration is to determine the key variables that can have negative effects of spaceflight and influence the crew. Although there is a body of research and knowledge on various negative aspects of spaceflight, including gravity or radiation, gained both in space and ground-based facilities together with some well-known medical counterparts, a better understanding of the psychological nature of long-term isolation is needed\cite{botella2016psychological}. This is crucial not only because of the above-mentioned changes in team composition compared to the Apollo program but also because the manned missions planned within the Artemis program and beyond, that is, for Moon and Mars, will last longer than previous Apollo missions (weeks and months/years vs. days). On the other side, they will also differ from previous long-term studies on orbital stations, namely the ISS, because of the different conditions such as Earth proximity or available space for the crew. Therefore, in line with the findings of ESA Topical Team Space Analogs and Human Performance, one of the fundamentals of our ALPHA-XR framework was established to determine key factors related to vital aspects of long-term manned missions, including sensory deprivation and social isolation. In particular, those factors can negatively affect executive and attentional functioning and therefore negatively impact skills, performance, efficiency, as well as well-being at both individual and crew level. 

To identify those factors, as well as methods to counteract negative effects, long-term observation and measurements are needed. Furthermore, to obtain the best results, research studies must be conducted under ecologically valid conditions, that is, a setting as similar as possible to daily routines in close proximity to the natural environment\cite{fonseca2023impact}. This means that the experimental study setup should reflect as much as possible the natural conditions of forthcoming long-term lunar manned missions. The best environment for conducting experimental studies is research laboratories. However, as we stated in our ALPHA-XR proposal, regular laboratory studies or training routines seem insufficient to induce proper levels of both social and individual factors in order to measure their interaction with, for example, executive and attentional functions that may influence individual and crew skills and performance as well as well-being. Therefore, to better prepare for real missions in space conditions, the research community is using special terrestrial facilities, called analog habitats. Some of them are part of official ground-based facilities of space agencies, but there are also a number of commercial sites that allow the organization of such simulated missions\cite{heinicke2021review}. Therefore, analog missions can serve as a useful experimental study vessel for the research community as a best approximation in terms of isolation factors studies. In addition to some aforementioned evident limitations, i.e. gravity or radiation, they provide an opportunity for conducting long-term isolation studies. 

\begin{figure*}[ht!]
 \centering
 \includegraphics[width=1\textwidth]{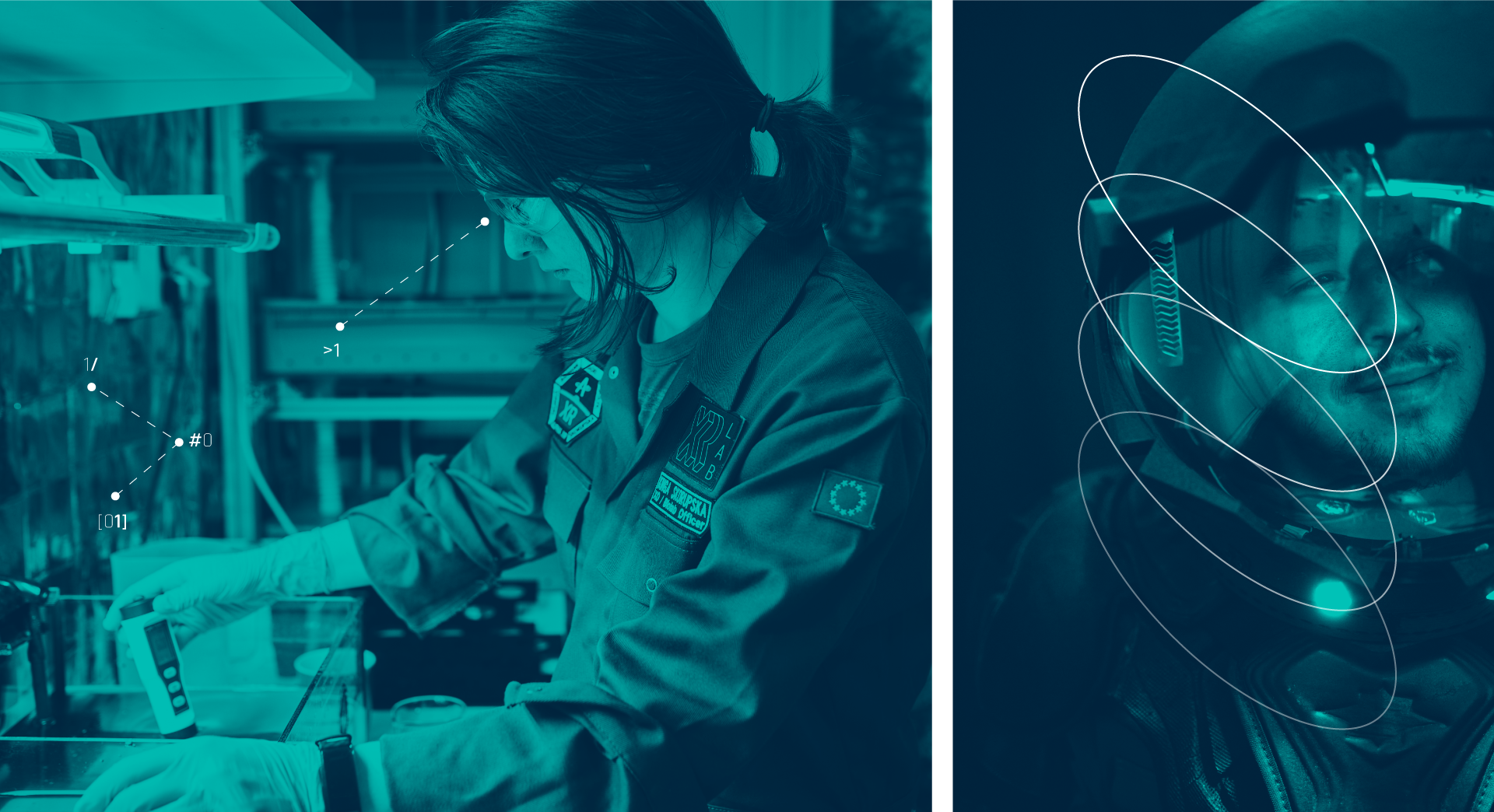}
 \caption{ALPHA-XR Simulated ICE Study crew routines. \\Photo by Suzanne Baumann, collage by Mateusz Szewczyk, Art\&Design Lab, XRC}
 \label{fig:alphaxrlab}
\end{figure*}

However, in regard to regular laboratory studies, such research activities, due to the nature of isolation and confinement, are to be conducted in an unsupervised approach. Among many limitations of such analog experiments, one prominent and fundamental restriction refers to methods and tools of measurement of various aspects of mental state of analog astronauts. In particular, as we mentioned, if we are interested in investigating participants' executive functioning, that is, their executive processes, like inhibition, updating of working memory, as well as attentional functions, we need to find proper ways of measurement in real-time, which is also in with the discussion of ESA Topical Team Space Analog and Human Performance that we contributed recently. The most established methods of measurement actually did not change for decades, besides moving from paper questionnaires towards their digital counterparts. In particular, those well-established self-reported tests, scales, and research protocols are not suitable for real-time measurement. Therefore, there is a chance to use new approaches derived from ecologically valid psychological laboratory studies, such as EMA (ecological momentary assessment), which involves repeated sampling of subjects' current behaviors and experiences in real time under natural conditions \cite{shiffman2008ecological,rosenkranz2020assessing}. However, based on our experience on the simulated mission, we know that the everyday routines of the crew, such as laboratory experiments or EVAs (extravehicular activities; fig. \ref{fig:alphaxrlab}), which are as important to provide ecological conditions as analog habitat environments, are so tedious and time-consuming that even the modern momentary assessment approach that we used (namely mobile applications instead of traditional paper forms) were prone to be omitted due to the crew's workload and responsibilities on site. Therefore, there is a need to further investigate the transition from active forms of measurements, that is, the forms that engage the crew, towards passive methods and tools, based on sensors and intelligent environment (IoT). However, this poses another challenge for the integration of multimodal data, including self-reported, psychological, and environmental or infrastructure-based measurements.

This leads to another research case study on IVR-based interventions during the analog mission\cite{ismar2023asthenia}, making active use of multimodal data gathering, including self-reported and psychophysiological data. Our case study proved that we may alleviate some negative factors, namely asthenia, towards increasing the well-being of the crew using artistic IVR experience. Based on our previous experience in such multimodal data-driven environments\cite{ismar2023metro,ismar2023vape} we conducted an experiment that unveiled the potential of artistic immersive experiences in favor of already well-established case studies on experiences based on nature exposed in VR. According to our findings, we can assert that artistic IVR experiences, particularly those that are interactive, demonstrate enhanced effectiveness in fostering well-being compared to traditional nature-based VR interventions. This enhancement suggests a unique potential for artistic IVR in impacting psychological outcomes, although the underlying mechanisms driving this effectiveness warrant further investigation. This is primarily attributed to the limitations inherent in our study, such as the small sample size characteristic of isolation studies in analog habitats. However, the use of XR-continuum-based solutions like IVR is still promising for overcoming the limitations of small-samples, one-shot studies, because the same setting may be used repeatedly in subsequent experiments in various locations, which is in line with our contribution to ESA Topical Team as well as outcome from the mentioned workshop on unsupervised condition\cite{interact2023kopec}.

Moreover, our IVR-based artistic intervention also unveiled some other shortcomings such as the complex and time-consuming process of creating such environments which also needs interdisciplinary collaboration. However, recent advances in generative AI also give some promising directions in this area for future research. 
Another interesting finding of ALPHA-XR is related to IVR-based systems application to the area of facilitating daily crew duties, namely towards effective communication between the crew and mission control \cite{skorupska2022case}, as well as some exploratory insights on rapid prototyping in terms of in-community experience of 3D printing in isolated, aiming for self-sufficiency\cite{stawski2023druk}. Moreover, based on our experience gained through the above research and activities in lunar habitat, we also organized follow-up activities, e.g space hackathon on participatory design of rapid prototyping solutions for the analog astronauts or even the future real space missions (fig. \ref{fig:xrhackathon}). Moreover thanks to the contacts and discussion in the broader circle of space researchers, including ESA EAC and ESTEC, another promising research area emerged. This was related to investigating XR-based immersive environments in lowering barriers towards space research facilities, including orbital equipment to broader community of researchers without previous experience in that area of conducting their research

To wrap up this year's panel discussion we would like to stress that exploring human factors in space exploration is an exciting yet challenging activity. Our exploratory research, based on extensive research of interdisciplinary cooperation in various fields, proves that there are some promising ways to overcome the current challenges in isolation and confinement simulated studies. However, they are also accompanied by specific challenges that yet may be overcome. This is the course of our current and planned research, that we hope to discuss at the next MIDI conference.

\section*{Acknowledgements}
The research leading to these results has received funding from the EEA Financial Mechanism 2014-2021 grant no. 2019\slash 35\slash HS6\slash 03166 (development of nature virtual environment).

%
%
%
\bibliographystyle{splncs04}
\bibliography{bibliography}

\end{document}